\documentclass{ws-procs9x6}

\usepackage{graphicx}% Include figure files
\usepackage{epsfig}
\usepackage{amssymb}

\begin{document}

\title{Precision Electroweak Constraints on Hidden Local 
Symmetries}

\author{R. Sekhar Chivukula, Elizabeth H. Simmons, \\
and Joseph
Howard\footnote{\uppercase{W}ork partially
supported by the \uppercase{US} \uppercase{D}epartment of \uppercase{E}nergy
under grant \uppercase{DE-FG02-91ER40676}.}}

\address{Physics Department\\
Boston University \\
590 Commonwealth Ave.\\ 
Boston, MA 02215 USA\\
E-mail: sekhar@bu.edu, simmons@bu.edu, and howard@bu.edu}

\author{Hong-Jian He}

\address{Center for Particle Physics and Department of Physics\\ 
University of Texas at Austin\\
Austin, TX 78712, USA\\
E-mail: hjhe@physics.utexas.edu}  


\maketitle

\abstracts{
In this talk we discuss the phenomenology of models with replicated
electroweak gauge symmetries, based on a framework with the
gauge structure $[SU(2){\ \rm or}\ U(1)]  \times  U(1) \times SU(2) \times  SU(2)$.}

\section{Generalized BESS}

In this talk we discuss the phenomenology of models with replicated
electroweak gauge symmetries. The general framework we use is based on
the gauge structure $[SU(2){\ \rm or}\ U(1)] 
\times  U(1) \times SU(2) \times  SU(2)$, and is
conveniently illustrated in the figure below. This figure is drawn
using ``moose''  notation,\cite{Georgi:1985hf} 
\begin{equation}
\lower20pt\hbox{\includegraphics[width=6cm]{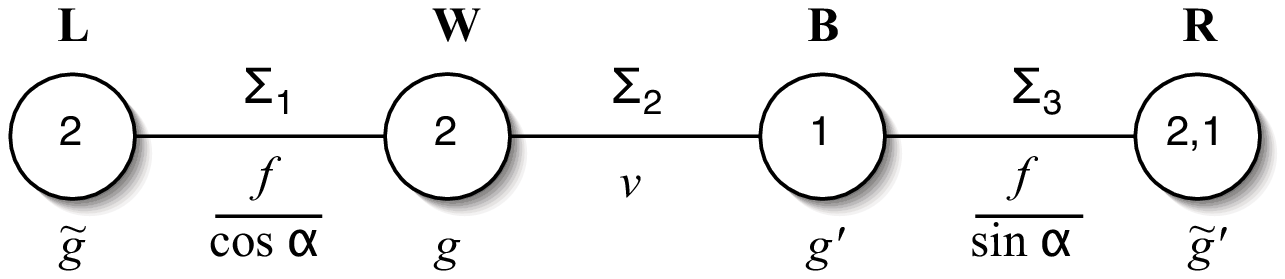}}
\end{equation}

in which the circles represent gauge
groups with the specified gauge coupling, and the solid lines
represent separate $(SU(2) \times SU(2)/SU(2))$ nonlinear sigma model
fields which break the gauged or global symmetries to which they are
attached. The solid circles represent $SU(2)$ groups, with a ``2''
denoting a gauged $SU(2)$ and the ``1''
a global SU(2) in which only a U(1) subgroup has been gauged.

For convenience, the coupling constants of the gauge theories will
be specified by
\begin{equation}
\tilde{g}^\prime = {e\over \cos\theta\, \sin\phi}\,,\ \ 
g^\prime = {e\over \cos\theta\, \cos\phi}\,,\ \ 
g= {e\over \sin\theta\, \cos\omega}\,,\ \
\tilde{g}={e\over \sin\theta\, \sin\omega}\,,
\end{equation}
and the $f$-constants (the analogs of $f_\pi$ in QCD) of the nonlinear sigma
models by
\begin{equation}
{f\over \sin\alpha}\,,\ \ v\,,\ \ {f\over \cos\alpha}~.
\end{equation}

As we will see, the Lagrangian parameters $e$, $\theta$, and $v$, will
be approximately equal to the electric charge, weak mixing angle, and
Higgs expectation value in the one-doublet standard model. The scale
$f$ sets the masses of the extra gauge bosons, and the theory reduces
to the standard model in the limit $f\to\infty$, while the angle
$\alpha$ allows us to independently vary the breaking of the
duplicated $SU(2)$ or $U(1)$ gauge symmetries. Finally, the angles
$\phi$ and $\omega$ determine the couplings of the gauge bosons which
become massive at scale $f$.

The symmetry structure of this model is similar to that proposed in the BESS
({\it Breaking Electroweak Symmetry Strongly})
model,\cite{Casalbuoni:1985kq,Casalbuoni:1995qt} an effective Lagrangian
description motivated by strong electroweak symmetry breaking.  This model is
in turn an application of ``hidden local symmetry" to electroweak
physics.\cite{Bando:1987br} Accordingly, we refer to this paradigm as
``generalized BESS." The symmetry structure in the limit $f\to\infty$ is
precisely that expected in a ``technicolor''
model,\cite{Weinberg:gm,Weinberg:bn} and the theory has a custodial symmetry
in the limit $g'$ and $\tilde{g}'$ go to zero.

Generalized BESS is the simplest model of an extended electroweak
gauge symmetry incorporating both replicated $SU(2)$ and $U(1)$ gauge
groups. As such the electroweak sector of a number of models in the literature form special cases, including {\it Noncommuting ETC},\cite{Chivukula:1994mn}
 {\it topcolor},\cite{Hill:1991at,Hill:1994hp}  and  {\it electroweak $SU(3)$}.\cite{Pisano:ee,Frampton:1992wt,Dimopoulos:2002mv}
The general properties of precision electroweak constraints on these 
models\cite{Chivukula:1995gu,Chivukula:1996cc,Csaki:2002bz} can correspondingly be viewed as special cases of what follows.\cite{Chivukula}

\section{Low-Energy Interactions}

Constraints on models with extended electroweak symmetries arise
both from low-energy and Z-pole measurements. The most sensitive
low-energy measurements arise from measurements of the muon
lifetime (which are used to determine $G_F$), atomic parity violation (APV),
and neutrino-nucleon scattering. In the usual fashion, we may summarize
the low-energy interactions in terms of four-fermion operators. The form
of these interactions will depend, however, on the fermion charge
assignments. For simplicity, in the remainder of this talk we
consider models in which the fermion charge assignments are flavor
universal. To illustrate the model-dependence of the results,
we consider two examples.

First, we consider the case in which the ordinary fermions are charged
only under the two groups at the middle of the moose
\begin{equation}
\lower25pt\hbox{\includegraphics[width=6cm]{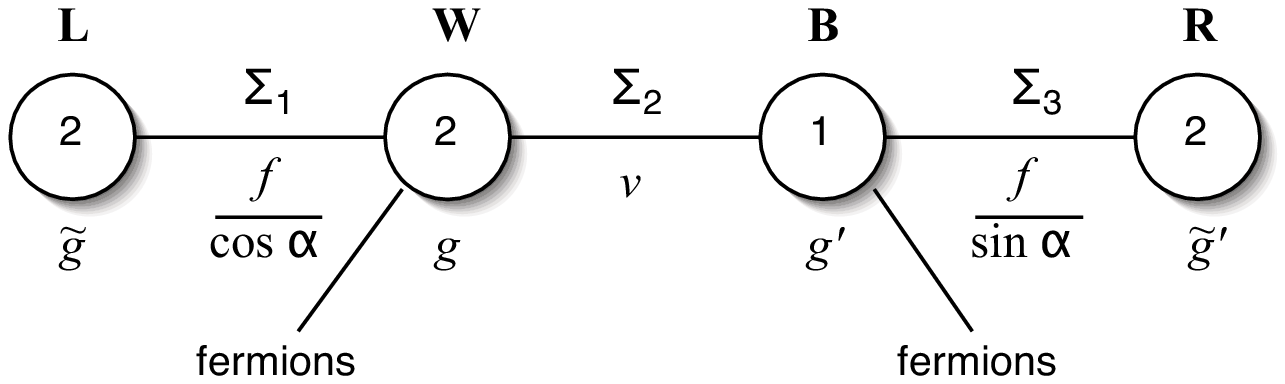}}\,.
\end{equation}
In this
case the charged current interactions may be computed to be
\begin{equation}
{L}^{CC} = -\,{2\over v^2}\, J^{\mu +} J^-_\mu\,,
\end{equation}
and the neutral current interactions
\begin{equation}
{L}^{NC} = -\,{2\over v^2}(J^\mu_3 - Q^\mu\, \sin^2\theta)^2 -\,
{2\cos^2\alpha \over f^2}\,\sin^2\theta\,\sin^4\omega\,Q^\mu Q_\mu~.
\end{equation}
In these expressions, the currents $J^\mu_{\pm,3}$ and $Q^\mu$
are the conventional weak and electromagnetic currents.
From these, we see that the strength of $G_F$, APV, and neutrino
scattering is determined by $v$ in the usual way. Furthermore, comparing
the two equations, we see that the strength of the charged and
neutral current interactions, the so-called low-energy $\rho$
parameter, is precisely one (at tree-level). This last fact
is a direct consequence of the Georgi-Weinberg neutral current
theorem.\cite{Georgi:1977wk}

As an alternative, consider the case in which the $SU(2)$ charges
of the ordinary fermions arise from transforming under the gauge
group at the end of the moose
\begin{equation}
\lower25pt\hbox{\includegraphics[width=6cm]{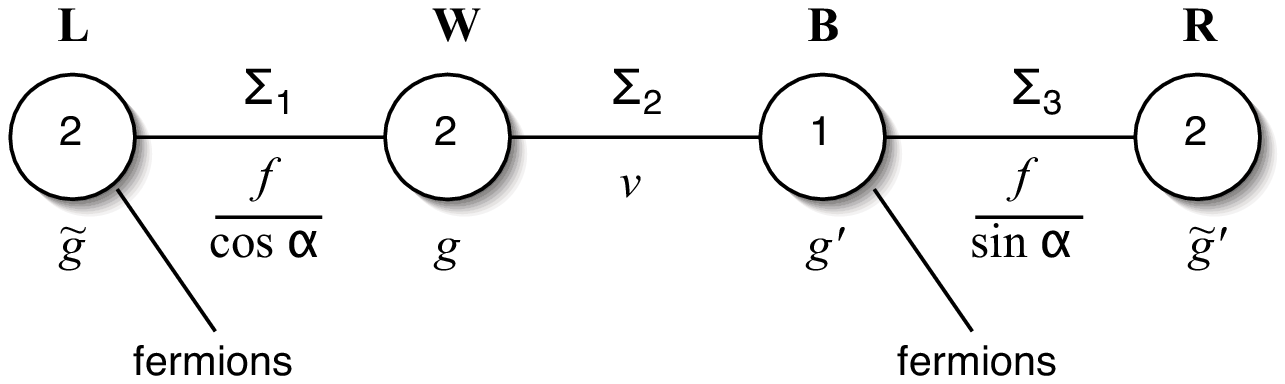}}\,.
\end{equation}
A calculation of the charged current
interactions yields
\begin{equation}
L^{CC}=-2\left({1\over v^2}+{\cos^2\alpha\over f^2}\right) J^{\mu +}J^-_\mu\,,
\end{equation}
while the neutral current interactions are summarized by
\begin{equation}
{L}^{NC} = -\,{2\over v^2}(J^\mu_3 - Q^\mu\, \sin^2\theta)^2 -\,
{2\cos^2\alpha \over f^2}\,(J^\mu_3-\sin^2\theta\cos^2\omega\,Q^\mu)^2~.
\end{equation}

Several points in this expression are of particular note: first, the value of $G_F$
as inferred from muon decay is no longer related simply to $v$. As
we shall see in the next section, this ultimately will give rise
to corrections to electroweak observables of order $(v/f)^2$
and unsuppressed by any ratios of coupling constants. Second,
unlike the previous case, the strength of low-energy charged-
and neutral-current interactions are no longer the same.
It is interesting to note, however, that the strengths of the $J_3^2$
and $J^+ J_-$ portions of the interactions are, however, the same -- 
this is a reflection of the approximate custodial symmetry
of the underlying model.

\section{Z-Pole Constraints - General Structure}

Many of the most significant constraints on physics beyond the
standard model arise from precise measurements at the Z-pole.  To
interpret these measurements, we must compute the masses $W$ and $Z$
bosons and their couplings to ordinary fermions in terms of the
Lagrangian parameters. For generalized BESS, we find the gauge-boson masses
\begin{equation}
M^2_W = {e^2 v^2\over 4\sin^2\theta}
\left(1-\cos^2\alpha \sin^4\omega\,{v^2\over f^2}\right)
+ O\left({v^4\over f^4}\right)~,
\end{equation}

and
\begin{equation}
M^2_Z = {e^2 v^2\over 4\sin^2\theta \cos^2\theta}
\left(1-(\cos^2\alpha \sin^4\omega+\sin^2\alpha\sin^4\omega)\,{v^2\over f^2}\right)
+\ldots
\end{equation}

From the expression for $M^2_Z$ and the calculations summarized in the
previous section, we immediately see that there is a major difference
in the structure of corrections to the standard model between cases I
and II: corrections to the standard model relation between $G_F$,
$\alpha$, $M^2_Z$, and the appropriately defined weak mixing angle
$\sin^2\theta_W$ are generically of order $v^2/f^2$ in case II, but is
of order $(\sin^4 \omega,\ \sin^4\phi) v^2/f^2$ in case I. As a
consequence, viewing the predictions of generalized BESS in terms of
{\it corrections} to the corresponding standard model results, the
corrections to standard model predictions in case I are (potentially)
suppressed by ratios of coupling constants relative to the size of
corrections in case II. This leads generically to
weaker constraints in case I models.

In what follows, we will concentrate on models in the category
of case I, in which the fermions are charged only under the gauge groups
in the ``middle'' of the moose diagram. 
In order to make predictions for electroweak observables,
we need to compute the couplings of the ordinary fermions to the light
gauge boson eigenstates. In the case of the $W$ we find
that the couplings to the left-handed fermions are
\begin{equation}
{e\over \sin\theta}\left(1-\cos^2\alpha\,\sin^4\omega\,{v^2\over f^2}\right)+\ldots
\end{equation}

and for the $Z$ we find the couplings
\begin{eqnarray}
{e\over \sin\theta\cos\theta} 
\left[ 1-  (\sin^2\alpha\,\sin^4\phi + \cos^2\alpha\,\sin^4\omega)\,
{v^2\over f^2}\right]\,T_3 \nonumber\\
 -\,{e\over \sin\theta\cos\theta} \left( \sin^2\theta-  \sin^2\alpha\,\sin^4\phi\,{v^2\over f^2}
\right)\,Q~.
\end{eqnarray}

Comparing to the computed gauge-boson masses we see that, for case I, all corrections to standard
model predictions may be expressed in terms of two combinations
of Lagrangian parameters:
\begin{equation}
c_1=\cos^2\alpha\,\sin^4\omega\,{v^2\over f^2}\,,\ \ \ \ \ 
c_2= \sin^2\alpha\,\sin^4\phi\,{v^2\over f^2}~.
\end{equation}
This allows us to compute bounds on model parameters in terms of fits
to $c_1$ and $c_2$, greatly simplifying the calculations.

Finally, while we will not explicitly display the results in case II,
a similar calculation shows that corrections to gauge-boson
couplings in this case are proportional to $(\sin^2\omega,\ \sin^2\phi)
v^2/f^2$.

\section{Flavor-Universal Results}

\begin{figure}[h]
\label{fig:Fig1}
\begin{center}
\includegraphics[width=6cm]{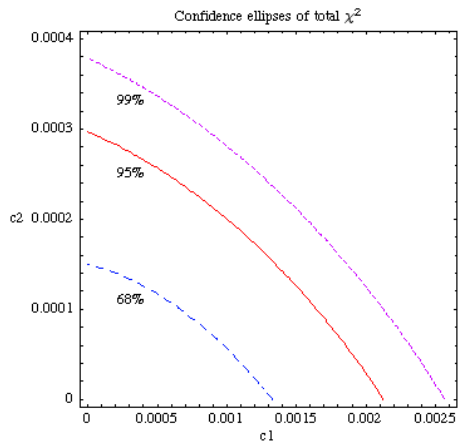} %\hspace*{8mm}
%\vspace*{-11mm}
\caption{
Constraints on $c_1$ and $c_2$ at the 68\%, 95\%, and
99\% confidence level based on fits to precision electroweak 
data.\protect\cite{Group:2002mc}
}
\end{center}
%\vspace*{-2.5mm}
\end{figure}

From the calculations above, we may compute the values of all
precisely measured electroweak quantities in terms of the Lagrangian
parameters given above. Using the procedure outlined
in Burgess {\it et. al.},\cite{Burgess:1993vc} we perform fits to the electroweak
observables listed in the most recent
compilation by the LEP Electroweak Working Group,\cite{Group:2002mc} which include Z-pole observables as well
as the width of the $W$ boson, and low-energy atomic parity violation
and neutrino-nucleon scattering. 
The 68\%, 95\%, and 99\% confidence level fits for the parameters
$c_{1,2}$ is shown in Figure 1. 

For a given value of
$\alpha$, we may unfold these constraints 
to produce a 95\% lower bound on $f$ in terms
of  $\sin\omega$ and $\sin\phi$. 
A sense of the reach of these
bounds is given in Figure 2, plotted for $\alpha=\pi/4$. For typical
values of $\sin\phi$ and $\sin\omega$, the bounds on the scale $f$ range from a few TeV. 

Many of the models cited above correspond
to the extra gauge groups being weak, $\sin\phi$ or $\sin\omega$ of order 1, 
in which case the bounds on $f$ are of order 10 TeV.\cite{Chivukula:1995gu,Chivukula:1996cc,Csaki:2002bz}
Formally the
corrections vanish when the couplings of the extra gauge groups
become strong, that is in the limit $\sin\phi\,, \sin\omega \to 0$. 
The phenomenologically interesting question is whether there
are {\it any} interesting models corresponding to this case,
in which case there may be interesting structure at relatively low scales!

\begin{figure}[h]
\label{fig:Fig2}
\begin{center}
\includegraphics[width=6cm]{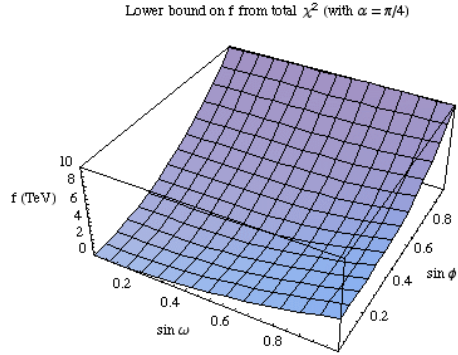} %\hspace*{8mm}
%\vspace*{-11mm}
\caption{
The lower-bound on $f$
at 95\% confidence level for $\alpha=\pi/4$, as a function of $\sin\omega$
and $\sin\phi$,  based on fits to precision electroweak 
data.\protect\cite{Group:2002mc}
}
\end{center}
%\vspace*{-2.5mm}
\end{figure}

\section*{Acknowledgments}

E.H.S. and R.S.C. thank to Koichi Yamawaki
and the rest of the organizing committee and staff for their hospitality and support.


\begin{thebibliography}{0}
%\cite{Georgi:1985hf}
\bibitem{Georgi:1985hf}
H.~Georgi,
%``A Tool Kit For Builders Of Composite Models,''
Nucl.\ Phys.\ B {\bf 266}, 274 (1986).
%%CITATION = NUPHA,B266,274;%%

%\cite{Casalbuoni:1985kq}
\bibitem{Casalbuoni:1985kq}
R.~Casalbuoni, S.~De Curtis, D.~Dominici and R.~Gatto,
%``Effective Weak Interaction Theory With Possible New Vector Resonance From A Strong Higgs Sector,''
Phys.\ Lett.\ B {\bf 155}, 95 (1985).
%%CITATION = PHLTA,B155,95;%%

%\cite{Casalbuoni:1995qt}
\bibitem{Casalbuoni:1995qt}
R.~Casalbuoni, A.~Deandrea, S.~De Curtis, D.~Dominici, R.~Gatto and M.~Grazzini,
%``Degenerate BESS Model: The possibility of a low energy strong electroweak sector,''
Phys.\ Rev.\ D {\bf 53}, 5201 (1996)
[arXiv:hep-ph/9510431].
%%CITATION = HEP-PH 9510431;%%

%\cite{Bando:1987br}
\bibitem{Bando:1987br}
M.~Bando, T.~Kugo and K.~Yamawaki,
%``Nonlinear Realization And Hidden Local Symmetries,''
Phys.\ Rept.\  {\bf 164}, 217 (1988).
%%CITATION = PRPLC,164,217;%%

%\cite{Weinberg:gm}
\bibitem{Weinberg:gm}
S.~Weinberg,
%``Implications Of Dynamical Symmetry Breaking,''
Phys.\ Rev.\ D {\bf 13}, 974 (1976).
%%CITATION = PHRVA,D13,974;%%

%\cite{Weinberg:bn}
\bibitem{Weinberg:bn}
S.~Weinberg,
%``Implications Of Dynamical Symmetry Breaking: An Addendum,''
Phys.\ Rev.\ D {\bf 19}, 1277 (1979).
%%CITATION = PHRVA,D19,1277;%%

%\cite{Chivukula:1994mn}
\bibitem{Chivukula:1994mn}
R.~S.~Chivukula, E.~H.~Simmons and J.~Terning,
%``A Heavy top quark and the Z b anti-b vertex in noncommuting extended technicolor,''
Phys.\ Lett.\ B {\bf 331}, 383 (1994)
[arXiv:hep-ph/9404209].
%%CITATION = HEP-PH 9404209;%%

%\cite{Hill:1991at}
\bibitem{Hill:1991at}
C.~T.~Hill,
%``Topcolor: Top quark condensation in a gauge extension of the standard model,''
Phys.\ Lett.\ B {\bf 266}, 419 (1991).
%%CITATION = PHLTA,B266,419;%%

%\cite{Hill:1994hp}
\bibitem{Hill:1994hp}
C.~T.~Hill,
%``Topcolor assisted technicolor,''
Phys.\ Lett.\ B {\bf 345}, 483 (1995)
[arXiv:hep-ph/9411426].
%%CITATION = HEP-PH 9411426;%%

%\cite{Pisano:ee}
\bibitem{Pisano:ee}
F.~Pisano and V.~Pleitez,
%``An SU(3) X U(1) Model For Electroweak Interactions,''
Phys.\ Rev.\ D {\bf 46}, 410 (1992)
[arXiv:hep-ph/9206242].
%%CITATION = HEP-PH 9206242;%%

%\cite{Frampton:1992wt}
\bibitem{Frampton:1992wt}
P.~H.~Frampton,
%``Chiral dilepton model and the flavor question,''
Phys.\ Rev.\ Lett.\  {\bf 69}, 2889 (1992).
%%CITATION = PRLTA,69,2889;%%

%\cite{Dimopoulos:2002mv}
\bibitem{Dimopoulos:2002mv}
S.~Dimopoulos and D.~E.~Kaplan,
%``The weak mixing angle from an SU(3) symmetry at a TeV,''
Phys.\ Lett.\ B {\bf 531}, 127 (2002)
[arXiv:hep-ph/0201148].
%%CITATION = HEP-PH 0201148;%%

%\cite{Chivukula:1995gu}
\bibitem{Chivukula:1995gu}
R.~S.~Chivukula, E.~H.~Simmons and J.~Terning,
%``Limits on noncommuting extended technicolor,''
Phys.\ Rev.\ D {\bf 53}, 5258 (1996)
[arXiv:hep-ph/9506427].
%%CITATION = HEP-PH 9506427;%%

%\cite{Chivukula:1996cc}
\bibitem{Chivukula:1996cc}
R.~S.~Chivukula and J.~Terning,
%``Precision electroweak constraints on top-color assisted technicolor,''
Phys.\ Lett.\ B {\bf 385}, 209 (1996)
[arXiv:hep-ph/9606233].
%%CITATION = HEP-PH 9606233;%%

%\cite{Csaki:2002bz}
\bibitem{Csaki:2002bz}
C.~Csaki, J.~Erlich, G.~D.~Kribs and J.~Terning,
%``Constraints on the SU(3) electroweak model,''
Phys.\ Rev.\ D {\bf 66}, 075008 (2002)
[arXiv:hep-ph/0204109].
%%CITATION = HEP-PH 0204109;%%

%\cite{Chivukula}
\bibitem{Chivukula}
R.~S.~Chivukula, H.~J.~He, J.~Howard, and E.~H.~Simmons, work
in progress.

%\cite{Georgi:1977wk}
\bibitem{Georgi:1977wk}
H.~Georgi and S.~Weinberg,
%``Neutral Currents In Expanded Gauge Theories,''
Phys.\ Rev.\ D {\bf 17}, 275 (1978).
%%CITATION = PHRVA,D17,275;%%

%\cite{Burgess:1993vc}
\bibitem{Burgess:1993vc}
C.~P.~Burgess, S.~Godfrey, H.~Konig, D.~London and I.~Maksymyk,
%``Model independent global constraints on new physics,''
Phys.\ Rev.\ D {\bf 49}, 6115 (1994)
[arXiv:hep-ph/9312291].
%%CITATION = HEP-PH 9312291;%%

%\cite{Group:2002mc}
\bibitem{Group:2002mc}
LEP Electroweak Working Group,
%``A combination of preliminary electroweak measurements and constraints  on the standard model,''
arXiv:hep-ex/0212036.
%%CITATION = HEP-EX 0212036;%%


\end{thebibliography}
\end{document}